\documentclass[aps,prc,reprint,showkeys,showpacs,nofootinbib,superscriptaddress,letterpaper,floatfix]{revtex4-1}
\usepackage{epsfig}  
\usepackage{amssymb}  
\usepackage{amsmath}  
\usepackage{amsfonts}  
\usepackage{color, graphicx}  
\usepackage{bm}
\usepackage[colorlinks=true,allcolors=blue]{hyperref}

\def\be{\begin{equation}}  
\def\ee{\end{equation}}

\def\Im{{\rm Im\,}}

\begin{document}  
 
 \title{Transmission coefficients in compound-nucleus reaction theory} 

 \author{Y.~Alhassid} \email{yoram.alhassid@yale.edu}
\affiliation{Center for Theoretical Physics,  Sloane Physics Laboratory,  Yale University, New Haven, Connecticut 06520, USA}
\author{G. F.~Bertsch} \email{bertsch@uw.edu}
\affiliation{Department of Physics and Institute for Nuclear Theory, Box 351560, University of Washington, Seattle, Washington 98195, USA}
\author{P.~Fanto} \email{paul.fanto@yale.edu}
\affiliation{Center for Theoretical Physics, Sloane Physics Laboratory,  Yale University, New Haven, Connecticut 06520, USA}
\author{T.~Kawano} \email{kawano@lanl.gov}
\affiliation{heoretical Division, Los Alamos National Laboratory, Los Alamos, NM 87545, USA}  

\begin{abstract} A recent article~\cite{br18} proposed a modification of the cross section
formula used in practical calculations of compound nucleus reactions.  We 
discuss the main concepts and approximations of statistical reaction theory and conclude that
the standard practical implementations of the cross section formula remain the preferred choice.
\end{abstract}  
  
  
\maketitle  
 
Although statistical reaction theory has played a prominent role in nuclear physics
since its beginnings in the 1930s, it has only come to maturity in
the 1980s with the development and analytic solution~\cite{we85,mi10} of the statistical model based 
on the Gaussian orthogonal ensemble (GOE) description of the underlying compound nucleus Hamiltonian.  
We shall refer to this model as the
GOE theory of compound nucleus reactions. 
The analytic solution for the average compound-nucleus reaction cross section in the GOE theory requires the evaluation of a complicated three-dimensional integral that depends on the parameters of the model,
and in practice simplified formulas are often implemented.  The most common such simplified treatment is the Hauser-Feshbach plus width-fluctuation-correction formula, discussed in detail below.

Ref.~\cite{br18} proposed a modification to this formula but did not test its performance with respect
to the underlying GOE theory.  In this work, we benchmark both the standard and modified formulas against the GOE theory.  We find that the standard formula is in good agreement with the GOE theory, while the modified formula deviates from the GOE theory at larger values of the coupling parameters. 

The average compound-nucleus reaction cross section $\bar \sigma_{c,c'}$ to scatter from channel $c$ to channel $c'$
can be expressed in terms of the fluctuating part of the $S$-matrix
\be\label{bar-sigma}
\bar \sigma_{c,c'} = \sigma_0 \overline{|\delta_{c,c'} -  S_{c,c'}|^2} \;,
\ee
where $\sigma_0 = \pi g_c/k_c^2 $ is the maximum inelastic cross section for incoming relative momentum $k_c$ in a two-body entrance channel $c$, and $g_c$ is a statistical spin factor in this channel.
We assume $N$ internal states described by an hermitean Hamiltonian $H$ coupled to $\Lambda$ open channels.  The coupling constants  $W_{\mu,c}$ between the internal states $\mu$ $(1 \le \mu \le N)$ and channels $c$ $(1 \le c \le \Lambda)$ form  an $N\times\Lambda$ real matrix $W$. The unitary $S$ matrix is expressed in terms of an hermitean $K$ matrix
\be\label{Smat}
S = \frac{1 - iK}{1+iK} \;,
\ee 
where the $K$ matrix is given by
\be
\label{Kmat}
K =  \pi W^T \frac{1}{E - H} W \;.
\ee
In the GOE theory, $H$ is chosen to be a random matrix that belongs to the GOE and the average over energy is replaced by an average over the ensemble. Expressed in the basis of eigenstates $\mu$ of $H$, 
the coupling  amplitudes $W_{\mu c}$ are characterized by channel-dependent Gaussian distributions in the limit of large $N$.
We define the partial width amplitude of an eigenstate
$\mu$ to decay into a channel $c$ by
\be
\gamma_{\mu c} = \sqrt{2 \pi} W_{\mu c} \;.
\label{eq:gamma_mu}
\ee  
A dimensionless coupling parameter $x_c$  in channel $c$ is defined by
\be
x_c = \frac{\pi}{2D} \langle \gamma_c^2\rangle \;,
\label{eq:xc}
\ee
where $D$ is the mean level spacing in the middle of the GOE spectrum, and $\langle \gamma_c^2\rangle$  is the mean square partial width amplitude in channel $c$
\be
\langle \gamma_c^2\rangle = \frac{1}{N} \sum_\mu \gamma_{\mu c}^2 \;.
\label{eq:gamma_square}
\ee

The ensemble-average $S$-matrix elements can be expressed in terms of $x_c$ (see Eq.~(65) in Sec.~V of Ref.~\cite{mi10})
\be\label{average-S}
\langle S_{c,c'} \rangle = \delta_{c,c'}\frac{1 - x_c}{1+x_c} \;.
\ee
The physical parameters are the transmission coefficients defined by
\be
T_c =  1 - |\langle S_{c,c}\rangle|^2\,.
\ee 
We note that in the presence of direct reactions, the average $S$ matrix is not diagonal and the transmission coefficient in channel $c$ is usually approximated by $T_c =  1 - \sum_{c'} |\langle S_{c,c'}\rangle|^2$. A rigorous method to treat the effect of direct reactions is to use the Engelbrecht Weidenm\"uller transform~\cite{en73}. 

\begin{figure}
\includegraphics[width=\columnwidth]{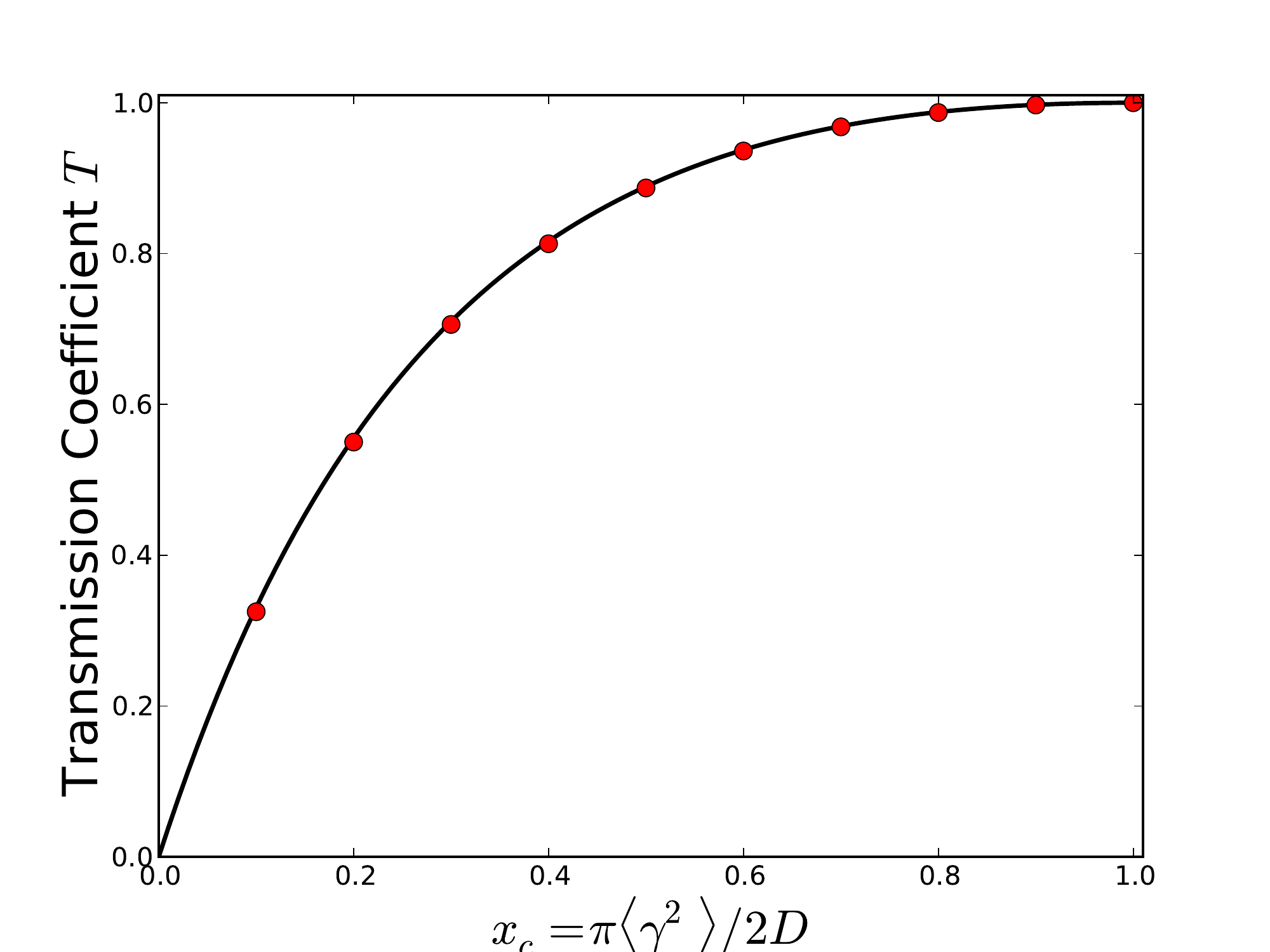}
\caption{
\label{tcf}
Transmission coefficient $T_c$ for an incoming channel $c$ in the GOE model with six equivalent reaction channels 
as a function of $x_c$.  
Solid line: Eq.~(\ref{x2T}).  Solid circles:
results obtained by sampling  $H$ from the GOE.
}
\end{figure}

Using Eq.~(\ref{average-S}), we have
\be
\label{x2T}
T_c = \frac{4 x_c} {(1+x_c)^2} \,.
\ee
Relations (\ref{average-S}) and (\ref{x2T}) hold for any values of the coupling parameters $x_c$.  
Thus the GOE theory is highly constrained; the complete structure of the 
average $S$-matrix elements at a fixed energy is 
determined by the the transmission coefficients $T_c$.  
Given the set of $T_c$,  the average compound-nucleus reaction cross sections
can be expressed in closed form by the three-dimensional
integral derived in Ref.~\cite{we85}.  Alternatively, 
these cross sections can be calculated numerically by sampling the Hamiltonian matrix $H$ from the GOE and using Eqs.~(\ref{Kmat}),  (\ref{Smat}), and (\ref{bar-sigma}) as was done in Refs.~\cite{ka14,ka15}.  This latter approach is demonstrated in Fig.~\ref{tcf}, which shows the incoming channel transmission coefficient $T_c$ for a reaction with six equivalent channels as a function of $x_c$. The numerical results obtained by sampling $H$ from the GOE (solid circles) are in excellent agreement with the analytic expression (\ref{x2T})  (solid line). 

However, the main practical codes implementing statistical reaction
theory follow the historical development of the theory
and compute compound-nucleus reaction cross sections in two steps
\cite{mo75}.  The first step is
to evaluate the Hauser-Feshbach (HF) formula for the average cross sections
or branching ratios, assuming that the 
relative decay probability in each channel is proportional to 
$T_c$ 
\be
\bar\sigma^{\rm HF}_{ab} = \sigma_0 \frac{T_a T _b}{\sum_cT_c} \;.
\ee
The average partial decay widths $\langle \Gamma_c \rangle$
may also be used as the input parameters (instead of $T_c$), but as discussed below, this
is more indirect approach. 
In the second step in constructing the practical formula, the HF cross section is corrected by a factor ${\cal W}_{ab}$, known as the width-fluctuation correction (WFC)
\be\label{CN}
\bar\sigma_{ab} = \bar\sigma_{ab}^{\rm HF}  {\cal W}_{ab} \;.
\ee
Since the complete GOE theory depends only on the parameters $x_c$ (or alternatively $T_c$), the WFC factor 
${\cal W}_{ab}$ can be calculated in terms of these parameters (e.g., using the triple integral formula of Ref.~\cite{we85}). 
A particularly accurate approximation to ${\cal W}_{ab}$ is Moldauer's 
formula~\cite{mo80} with a particular parametrization of the number of degrees of freedom  $\nu_c$ in channel $c$ in terms of the transmission coefficients (see, e.g., Ref.~\cite{ka14}).
   
In Ref.~\cite{br18}, the average partial widths of the resonances 
$\langle\Gamma_c \rangle$ were taken as the parameters of the statistical
theory.  In principle, these partial widths can be calculated from the pole expansion of the $S$ matrix, and the total width can be determined by the imaginary part of the
$S$-matrix poles $E_\mu$ in the complex energy plane,
$\Gamma_{\mu} = -\Im E_\mu/2$.   In the weak-coupling limit where $x_c$ are small in all channels, the relation between $\langle\Gamma_c \rangle$ and $x_c$ is
\be
\label{x2G}
\frac{\langle\Gamma_c\rangle}{D} \approx \frac{2}{\pi} x_c\;.
\ee
As discussed in Ref.~\cite{br18}, average partial widths can be measured in the unresolved resonance region where Eq.~(\ref{x2G}) no longer holds. 
It would be useful to relate the average partial widths to the transmission coefficients in order to test optical model parameterizations of the latter.
This will be addressed in the framework of the GOE theory in future work. Here we use the parameters $x_c$  defined by Eqs.~(\ref{eq:gamma_mu}) and (\ref{eq:gamma_square}), or alternatively $T_c$, as the independent parameters of the statistical theory. 

Ref.~\cite{br18} proposed a modification of the HF+WFC formula (\ref{CN}) by
replacing $T_c$ in Moldauer's formula by a certain function $f_c$ of $T_c$.  The proposed 
formula has the structure
\be
\label{mod}
\bar\sigma^{\rm mod}_{ab} = \sigma_0\frac{T_a f_b}{\sum_c f_c}{\cal W}_{ab}(\vec f) \;.
\ee
Here $\vec f= (f_1,\ldots,f_\Lambda)$ and 
\be
\label{calW}
{\cal W}_{ab}(\vec f)
= \left(1 + \delta_{ab}\frac{2}{\nu_a}\right) \int_0^\infty dz \prod_c\left( 1+ \frac{2 f_c z}
{\nu_c\sum_{c'} f_c'}
\right)^{-\delta_{ac}-\delta_{bc} -\nu_c/2} \;,
\ee
where $\nu_c$ is the number of degrees of freedom in channel $c$. 
In particular, the authors of Ref.~\cite{br18} find that
the choice 
\be
\label{Tbrown}
f_c  = \frac{T_c}{\sqrt{1 - T_c}}
\ee
permits a better agreement with experiment in one of the reactions they
studied.

To determine whether this modified formula provides better fits to
cross sections of the statistical GOE theory when the coupling parameters $x_c$ vary between the 
weak coupling limit $x_c \ll 1$ and $x_c=1$, we consider a two-channel compound nucleus reaction with similar parameters to 
Example C of Ref.~\cite{ka14}. In our
calculation we use the same definition of the WFC as in Refs.~\cite{ka14,br18}.

We first vary the transmission coefficient $T_a$ in one channel from zero to one,
setting $T_b=0.26$ in the second channel.  The results are shown
in Fig.~2a, where the ratio of the cross section computed using $f_c$ in 
Eq.~(\ref{Tbrown}) to the cross section calculated in the GOE theory is compared with a similar ratio but using the standard formula for the WFC in which $f_c= T_c$.
We observe that this ratio is close to $1$ for the standard choice of variables
$f_c = T_c$ (solid black circles).  This is not surprising since the particular parameterization of
the WFC was obtained by fitting such ratios.  If we use $f_c$ as in
 Eq.~(\ref{Tbrown}), we obtain the higher cross section ratios shown in solid red circles, indicating deviations from the GOE theory as $x_a$ increases from its value at $x_a=0.1$. 
 
 \begin{figure}
\includegraphics[width=\columnwidth]{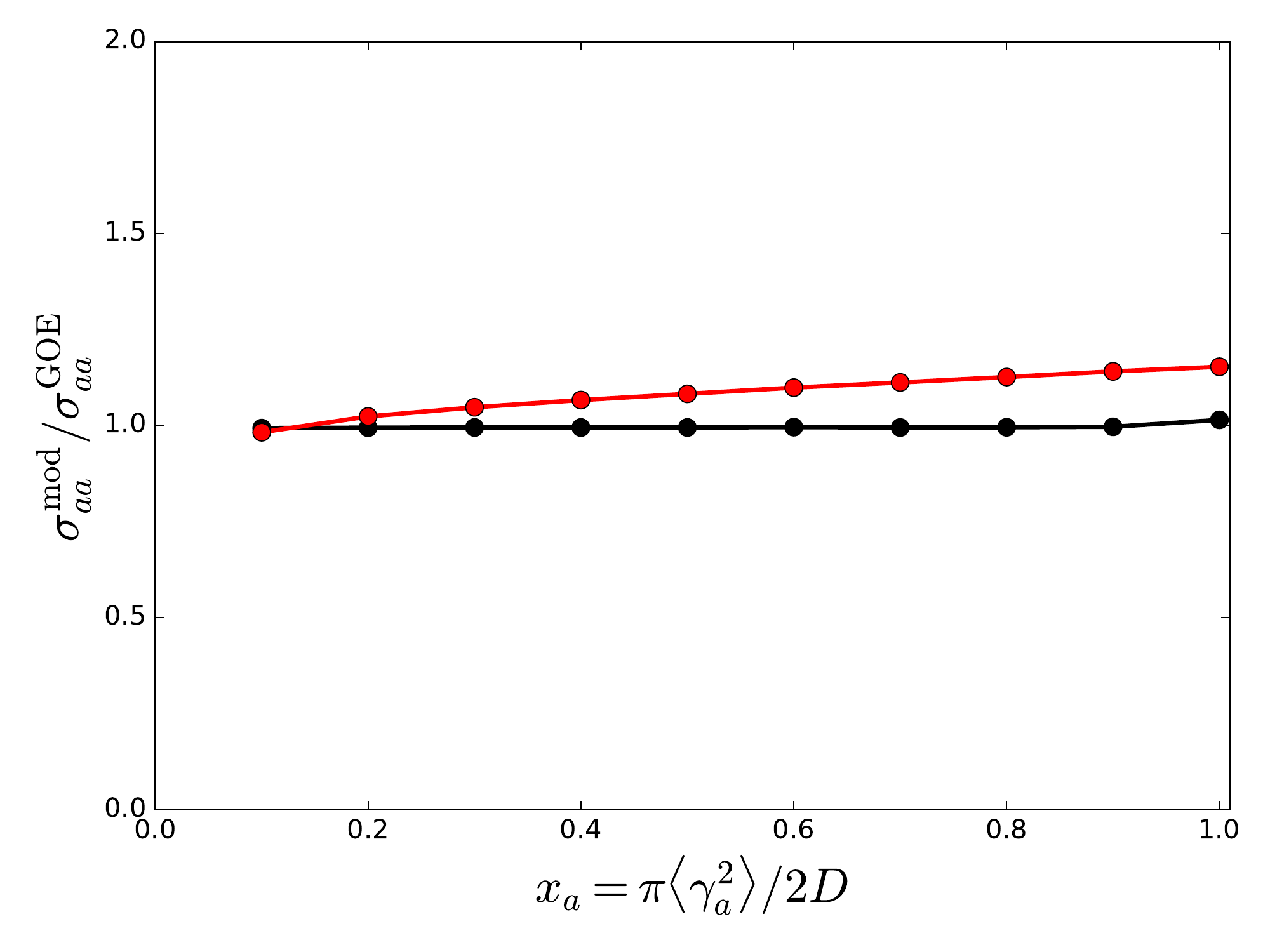}
\includegraphics[width=\columnwidth]{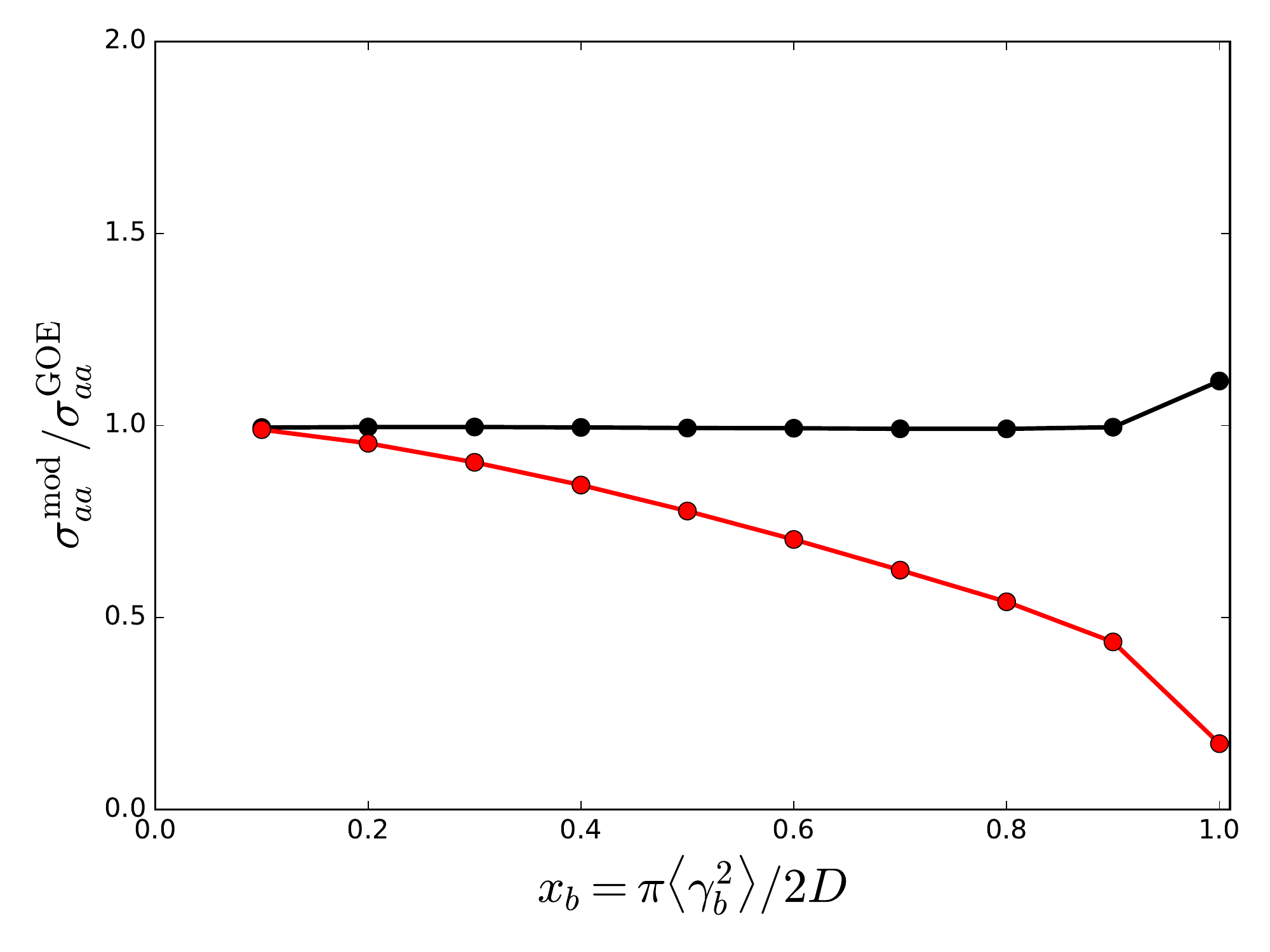}
\caption{
\label{sigma} 
Ratio of entrance channel cross section calculated from Eqs.~(\ref{mod}) and (\ref{calW})  to the GOE statistical theory cross section
for the two-channel example described in the text.  Black circles:
$f_c = T_c$;  Red circles:  
$f_c = T_c/\sqrt{1-T_c}$.  Panels (a) and (b) show the ratios as a function
of $x_a$ and $x_b$, respectively.
}
\end{figure} 
  
 Next, we make a similar comparison by varying $T_b$ while setting  $T_a=0.39$. The results are shown in Fig.~2b.  Again, the standard choice $f_c=T_c$ gives very good agreement with the GOE theory, while the choice in Eq.~(\ref{Tbrown}) leads a reduction in the ratio at higher values of $x_b$.  Thus choosing a different parametrization for $f_c$ does not improve the fits without making a
compensating change in the parametrization of the WFC factor $\cal W$.

We close with some remarks on the superradiance phenomenon.
In statistical reaction theory, superradiance refers to the development of broad resonances with simple structure for  $x_c > 1$. 
However, these resonances do not affect the average cross sections of the statistical theory~\cite{mi10}.  
From Eq.~(\ref{x2T}), we see that the transmission coefficients $T_c$ (\ref{x2T}) are invariant the transformation $x_c \to 1/x_c$.  Since the analytic solution of the GOE theory depends only on $T_c$, changing from $x_c < 1$ to $1/x_c$ at which superradiance occurs does not change the GOE theory cross section.
 
In conclusion, we advocate using the statistical GOE
theory as far a possible in analyzing compound-nucleus reaction cross sections.

\section*{Acknowledgments}
The work of Y.A. and P.F. was supported by the U.S. DOE grant
Nos.~DE-FG02-91ER40608 and by the U.S. DOE 
NNSA Stewardship Science Graduate Fellowship under cooperative agreement 
No.~NA-0003864.
T.K. carried out this work under the auspices of the
National Nuclear Security Administration of the U.S. Department of Energy at
Los
Alamos National Laboratory under Contract No. DE-AC52-06NA25396.

\end{document}